# Computation of gray-level co-occurrence matrix based on CUDA and its optimization

Huichao Hong, Lixin Zheng, Shuwan Pan
Engineering Research Center of Industrial Intelligent Technology and Systems of Fujian Providence
College of Engineering,Huaqiao University,Quanzhou, China
e-mail: zlxgxy@qq.com

Abstract: As in various fields like scientific research and industrial application, the computation time optimization is becoming a task that is of increasing importance because of its highly parallel architecture. The graphics processing unit is regarded as a powerful engine for application programs that demand fairly high computation capabilities. Based on this, an algorithm was introduced in this paper to optimize the method used to compute the gray-level co-occurrence matrix (GLCM) of an image, and strategies (e.g., "copying", "image partitioning", etc.) were proposed to optimize the parallel algorithm. Results indicate that without losing the computational accuracy, the speed-up ratio of the GLCM computation of images with different resolutions by GPU by the use of CUDA was 50 times faster than that of the GLCM computation by CPU, which manifested significantly improved performance.

With the increasing data size and complexity of image, the fast processing of it in the course of actual use is a problem that necessitates solutions. The required processing time is usually limited to be within a reasonable period. Therefore, numerous scholars have made much effort to design high-performance architectures suitable for image processing. This paper aims to introduce a method that we developed based on CUDA, used to accelerate the texture-based computation of the image characteristics.

Texture refers to the repetition of a certain type of structure or pattern in some image areas on the real objects, meaning a texture image consists of many identical or similar elements. These elements are called primitives. Therefore, the texture image analysis is to find a suitable method to extract the primitive characteristics or the characteristics of areas formed by primitives. [1] The GLCM is one of the most famous second-order statistical characteristics analysis methods for texture. To estimate similarities between different gray-level co-occurrence matrices (GLCMs), Haralick proposed extracting fourteen statistical characteristics from each GLCM. [2] When processing high-resolution images, such as an image of pathological tissue section that has cancer texture and normal texture in different areas, we usually, in these cases, compute the GLCMs of different frequency bands in the same target based on the multispectral image, resulting in the evident rise of the computation amount. [3-5] Related work was also reported in remoting sensing. [6-7]. However, computers that are limited by their adoption of the von Neumann architecture have failed to use the parallel computation that is inherent in the GLCM. [8] Therefore, the highly parallel architecture is needed for parallel realization.

In recent years, general-purpose computing on graphics processing units (GPGPU) has become a successful trend of high-performance computation under programming environment such as CUDA [8] and OpenCL. [9] A large quantity of parallel units are available that operate on hardware in the multi-thread way. Application programs based on the large-scale paralleling, benefiting from the trend, have seen impressive growth.

This study has three major contributions:

(1)A method that computes GLCM based on CUDA was proposed, which saw great improvement compared to the computation on the CPU platform.

(2)The reason why the gray level and the changes of gray level affect the GPU computation time was discussed, which was mainly because the read-write collisions occurred when multi threads were reading and writing the same address, resulting in serialization of the threading operation. For the images with slow gray level changes, because of the high correlation of elements in the neighboring areas, the read-write collisions was more easily attributed to the changes of the gray level instead of the gray level. For the images with drastic gray level changes, because of the low correlation of elements in the neighboring areas, the frequency of operation that generates atoms was less than that of the image data, the gray level had more impacts than the image data did for this type of data. Because of the similar principles, analyses above serve as a reference to the analysis of the image statistical histogram.

(3)We came up with a solution for the problems above, i.e., putting multiple GLCM copies in the shared memory of each active block. Threads in the block will write the voting results into different copies under certain rules. Details are amplified in Section 2 below.

Because data transmission from the Host to the Device takes half of the total time, the efficiency and the timeliness of the program operation are greatly restricted. To solve the problem, this paper proposed a CUDA-stream-based strategy of processing the image by block.

## I. GLCM and analysis of parallelism

### A. GLCM

GLCM is an Nth-order matrix used to describe the joint distribution probabilities of pixel pairs, where N is defined as the image gray level. To reduce the computing complexity and highlight the texture characteristics, the image gray level will usually be lowered to 8, 16 or 32 at the stage of pre-processing. [2]

The mathematical definition of GLCM: frequencies ( $\mathbf{P}(i,j;d,\theta)$ ) of simultaneous occurrence of pixel $(x_1,y_1)$ and pixel $(x_2,y_2)$. Distance between pixel $(x_1,y_1)$ and pixel $(x_2,y_2)$ is d and pixel $(x_2,y_2)$ is to the $\theta$ direction of pixel $(x_1,y_1)$.

$$\mathbf{P}(i,j;d,\theta) = \#\{(x_1,y_1)(x_2,y_2) \mid f(x_1,y_1)=i, f(x_2,y_2)=j, \mid (x_1,y_1) - P(x_2,y_2) \mid = d, \angle((x_1,y_1),(x_2,y_2)) = \theta\} \quad (1)$$

### B. Design and analysis of GLCM parallel program (Scheme 1)

The platform in this study is Intel(R) Core(TM) i5-4590 with a clock rate of 3.6GHz, working with 8GB RAM and a graphics card that is NVIDIA GeForce GTX 1050Ti, of which the specifications are listed in Table I.

**Table I. Specifications of NVIDIA GeForce GTX 1050Ti**

| Architecture | CUDA core unit | Video memory |
|---|---|---|
| Pascal | 768 | 4096MB |
| Memory interface | GPU clock | Memory clock |
| 128bit | 1354/1468MHz | 7000MHz |

All experiments in this paper were conducted in Windows7 SP1 with the development environment of VisualStudio2013 Update5. The adopted CUDA modules included N-sight, CUDA 8.061. The driver of the graphics card was 378.66.

The main steps of establishing a program based on the CUDA parallel mode were as follows: memory was assigned in the host to store the image data read from the external device and the results from the GPU computation (i.e., GLCM, the same hereinafter). Global memory was assigned in the device to receive the image data from the host, and store and return the results. Secondly, the host called the kernel function to conduct the computation in GPU. Finally, when the results met the discontinue rule, the GLCM from the computation was returned and the global memory in the device for the storage of the inbound image data was released.

The core idea of the GLCM computation based on GPU lies in how to ensure all pixel pairs that fit Formula (1) can be processed in parallel and independently.

For the image $f(x,y)$, where $x \in (1,N)$ and $y \in (1,N)$, set the gray level at pixel $(x\_associate, y\_associate)$ as $f\_associate$, and $f\_ref$ as the gray level at pixel $(x\_ref, y\_ref)$. Distance between the two pixels is $d$ and direction of one towards the other (direction along with GLCM is computed) is $\theta$.

Positional relation between $f\_associate$ and $f\_ref$ is as shown in Formula (2), provided that the computer applies the row main sequence storage:

$$f\_ref\_addr = \begin{cases} f\_associate\_addr + d, & \theta = 0° \\ f\_associate\_addr + d \times (N-1) & \theta = 45° \\ f\_associate\_addr + d \times N, & \theta = 90° \\ f\_associate\_addr + d \times (N+1), & \theta = 135° \end{cases} \quad (2)$$

A piece of global memory needs to be initialized to store GLCM [1] $\mathbf{P}[1...l^2]$, where $l$ refers to the gray level of image. The position coordinates $pos$ of pixels in this global memory, the gray level $f\_associate\_val$ of pixel $f\_associate$, and the gray level $f\_ref\_val$ of pixel $f\_ref$ are as shown in Formula (3):

$$pos = f\_ref\_val \times l + f\_associate\_val \quad (3)$$

Obtain the pixel position coordinates in $\mathbf{P}$ by the use of the gray level $f\_associate\_val$ of pixel $f\_associate$, and the gray level $f\_ref\_val$ of pixel $f\_ref$ based on Formula (3), then add 1 to both values of X axis and Y axis in the coordinates (such operation can be called voting). After reading and voting of $f\_ref\_val$ and the neighboring $f\_associate\_val$ of each pixel, the obtained $\mathbf{P}[1...l^2]$ is equal to $\mathbf{P}(i,j;d,\theta)$.

According to the process above, the time complexity of the algorithm is $O(N^2)$, which is fairly high for real-time application.

However, it can be easily seen that each pixel is processed independently from the reading of its $f\_associate\_val$, $\theta$, $d$, the gray level $f\_associate\_val$ of its neighboring pixel, to adding 1 to the pixel coordinates in $P$ consisted of $f\_associate\_val$ and $f\_ref$. Therefore, simultaneous computation of each pixel can be achieved by the use of the numerous threads in NVIDIA CUDA.

It should be pointed out that despite the independent processing, thread conflict will occur if multiple CUDA threads read and write the same position. Therefore, the atomic operation is needed to ensure every time a latest value is read. By the addition of the atomic operation, the maximum number of threads operating simultaneously is the square of the gray level $L$ and the rest of the threads will be lining up. Performance analysis and running time are provided in detail in Section 2.

## II. Optimization of GLCM computation based on copy mechanism (Scheme 2)

### A. Performance analysis and method introduction

Multiple CUDA threads' Reading and writing of the same position of GLCM causes memory conflict. Memory conflict is closely related to the gray level and its distribution across the image.

Images in Fig.1 were selected as the sample for the experiment. As shown in Fig.1(a) and Fig.1(b), magnitudes of the gray level changes vary, with (a) being small and (b) being significant. Their respective gray level was quantized to 8 and 32 in consideration of the impacts of the gray level on the algorithm efficiency.

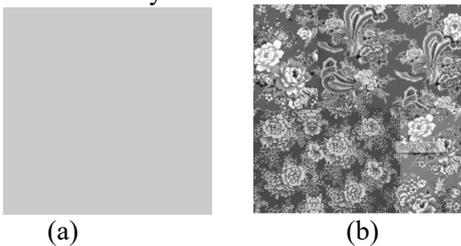

(a)                (b)
Fig.1 Images with different gray level changes

The algorithm ran following the steps indicated in Scheme 1 (Section 1.2). The test time was defined as the average duration of 20 tests (only kernel runtime was measured). Size of all images was 1024 × 1024. Included angles between pixels were 0° and 45° respectively, with their own corresponding distance of 1 and 4. Experimental data are shown in Table II.

**Table II.**

| Image | Gray level | Running time/ms | | | |
|---|---|---|---|---|---|
| | | $d=1$, $\theta=0°$ | $d=1$, $\theta=45°$ | $d=4$, $\theta=0°$ | $d=4$, $\theta=45°$ |
| Fig.1(a) | 8 | 0.96 | 0.92 | 0.91 | 0.91 |
| | 32 | 0.90 | 0.89 | 0.90 | 0.90 |
| Fig.1(b) | 8 | 0.92 | 0.92 | 0.94 | 0.91 |
| | 32 | 0.28 | 0.30 | 0.27 | 0.27 |

For the data illustrated, we assume that the number of threads that GPU could deploy to process the pixel pairs and add 1 to values of the position coordinates formed by pixels was $N_p$.

The conclusion can be drawn by comparison of the experimental data yielded from the same gray level of Fig.1(a) and Fig.1(b):

(1) When the image moves towards a certain direction and the neighboring gray levels vary greatly, those $N_p$ threads will read and write scattered areas in GLCM because of the weak correlation between two pixels. Consequently, probabilities that different threads access the same memory address simultaneously will reduce. (2) When the image moves towards a certain direction and the neighboring gray levels change little, those $N_p$ threads will focus on reading and writing some addresses, thus increasing probabilities that different threads access the same address simultaneously.

Conclusion can be drawn according to the experimental data from different gray levels in Fig.1(b):

As stated in Section 1.2, the GLCM size is the square of the image gray level $L$. When the size gets bigger, threads will visit areas that separate from each other. Hence, probabilities of different threads accessing the same address simultaneously will lower.

As for these problems, we proposed a GLCM optimization method that applies a strategy of using multiple copies. It aims to lower probabilities of multiple threads reading and writing the same address simultaneously. Its specific principles and implementation are described below:

The first level cache and shared memory in each stream multiprocessor has a 64K memory segment. According to program requirement, each active block in the stream multiprocessor has a private memory that has the same size but no more

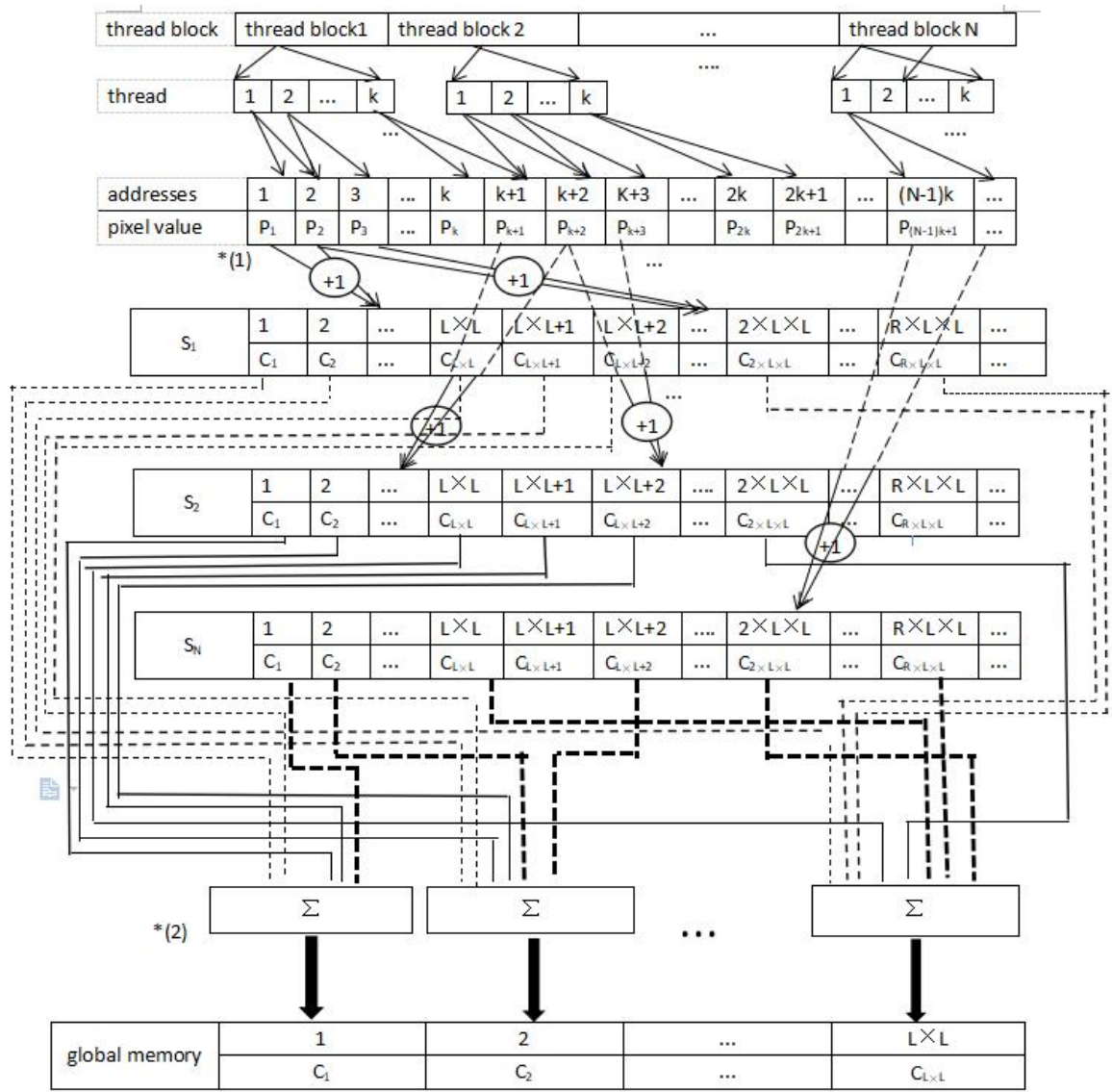

Fig.2 Diagram of GLCM computation optimization method based on copy strategy
(when θ=0°, d=1)

than 48K, meaning this private piece of memory can only be accessed by threads in the block. [2]

Based on this character, each block (the number of thread included is k) only processes a certain area of image. Meanwhile, in each block's share memory, we established a total of R continuous regions with each size of $L \times L$, for the storage of GLCMs (sub-GLCMs hereafter).

Furthermore, the i-th thread in the block voted to the position ($P+Q\times L$) in the (i MOD R)-th sub-GLCM, where P and Q were the corresponding pixel values in the i-th block, and L was the gray level.

As stated in Section 2, pixels between neighboring areas tend to be highly correlated. Hence, even for the identical pixel values in the neighboring areas, the conflict magnitude can be reduced by R times after the steps taken above.

Particularly, the operation symbol "*1" was seen identical to the voting process described in Section 1.2.

Because each block had a shared memory and R sub-GLCMs within, there were $R \times N$ (N is the number of blocks) sub-GLCMs in total. The final result was the sum of pixel values in all sub-GLCMs. The result was stored in the global memory.

B. Experiment of GLCM computation optimization method based on copy strategy and analysis

Proper grid and block dimension design can give rein to the GPU potency. In general, the larger the grid dimension the better, while in the block dimension design, factors below should be considered [11]:

Multiple active blocks are assigned by the stream multiprocessor. These active blocks were supposed to operate in sequence and asynchronous execution of both accessing and computation was available. When one block was doing the high-latency job such as syncing or visiting the video memory, the other block could occupy the GPU resources. The application of multiple active blocks was able to conceal the visiting latency. Therefore, the principles of setting the number of active blocks in the stream multiprocessor should be

$$\frac{ThreadMaxInSM}{b_{dim}} \geq n_b \geq 2 \quad (4)$$

Meanwhile, the product of the active block number, the sub-GLCM number in each active block (R, R≥1) and the sub-GLCM size (S) should be no larger than the max shared memory that can be assigned by the GPU block (SharedMemoryMaxInSM, usually 48k):

$$\frac{SharedMemoryMaxInSM}{R \times S} \geq n_b \quad (5)$$

Based on the limitation above, for different gray levels, R should meet the following formula:

$$\frac{SharedMemoryInSM}{n_b \times S} \geq R \geq 1 \quad (6)$$

According to Formula (5), for the gray levels of 32 and 8, R should be no more than $\frac{12}{n_b}$ and $\frac{48}{n_b}$, respectively.

After a number of experiments, we concluded that under the Formula (5) and (6):

Table III. Runtimes under different resolutions, gray levels and GLCM parameters by algorithm in Scheme 2

| Gray level | image size | Image | Runtime(/ms) | | | |
|---|---|---|---|---|---|---|
| | | | d=1, θ=0° | d=1, θ=45° | d=4, θ=0° | d=4, θ=45° |
| 8 | 1024×1024 | Fig1.(a) | 0.37 | 0.38 | 0.38 | 0.37 |
| | | Fig1.(b) | 0.40 | 0.34 | 0.35 | 0.44 |
| | 4096×4096 | Fig1.(a) | 2.40 | 2.43 | 2.44 | 2.47 |
| | | Fig1.(b) | 2.87 | 2.49 | 2.40 | 2.46 |
| | 8192×8192 | Fig1.(a) | 8.85 | 9.53 | 9.07 | 9.33 |
| | | Fig1.(b) | 8.93 | 9.35 | 8.78 | 9.45 |
| | 16384×16394 | Fig1.(a) | 34.97 | 36.93 | 34.76 | 37.60 |
| | | Fig1.(b) | 34.95 | 36.89 | 35.06 | 37.47 |
| 32 | 1024×1024 | Fig1.(a) | 0.40 | 0.41 | 0.40 | 0.39 |
| | | Fig1.(b) | 0.38 | 0.39 | 0.37 | 0.39 |
| | 4096×4096 | Fig1.(a) | 2.51 | 2.56 | 2.56 | 2.53 |
| | | Fig1.(b) | 2.93 | 2.52 | 2.46 | 2.83 |
| | 8196×8196 | Fig1.(a) | 9.26 | 9.26 | 9.30 | 9.62 |
| | | Fig1.(b) | 9.16 | 9.35 | 9.01 | 9.52 |
| | 16384×16394 | Fig1.(a) | 36.09 | 37.72 | 36.35 | 37.98 |
| | | Fig1.(b) | 36.05 | 37.48 | 36.01 | 38.15 |

- For the gray level of 32 and R no less than 2, the highest efficiency occurred when the block size was 512.
- For the gray level of 8 and the block size of 1024, the highest efficiency occurred when R was no less than 4.

Images of different sizes (1024×1024, 2048×2048, 4096×4096 and 8196×8196) were chosen to measure the runtimes in Scheme 2.

As shown in Table III, compared with the results in Scheme 3, performances in Scheme 2 were poorer by 100% except for the condition where the target was Fig.1(b) and the gray level was 32 (reason was discussed in Section 2. It not only fully indicated the effectiveness of the scheme proposed and demonstrated our point of view described in Section 2, i.e., the atom conflict is one of the major factors that impede the GLCM computation.

## III. Strategy of processing image by block based on CUDA streams(Scheme 3)

In Section 2.1, the algorithm execution follows the main steps as follows:
- Memory is assigned in the host for the storage of the image data read from the external device and the computation results returned by GPU

- (GLCM hereafter).
- Global memory is assigned in the device to store the image data, and save and return the computation results from the host.

Based on Scheme 2, images of different sizes (1024×1024, 4096×4096, 8196×8196 and 16384×16384) were chosen. For example (when d=1 and θ=0°), the time of transmission from host to device and from device to host was measured, and so was the GPU kernel runtime (a, b and c, respectively).

Table 3 Time of transmission and computation of GLCM under d=1 and θ=0°

| Image size | Runtime/ms | |
| --- | --- | --- |
| | Transmission time | Computation time |
| 1024×1024 | 0.25 | 0.15 |
| 4096×4096 | 2.01 | 0.86 |
| 8196×8196 | 5.89 | 3.03 |
| 16384×16384 | 22.99 | 11.96 |

exeStream (for kernel codes execution) and copyStream (for copying data from the host to the device). When exeStream was processing the first input image data in the buffer zone, copyStream started to copy the second image to the second buffer zone simultaneously. Once exeStream finished processing, the GPU kernel would initiate again. [10]

Based on the principles above, a strategy of processing image by block based on CUDA streams was proposed – transmitting and processing image by block.

Two streams were defined in the program. Each stream executes the same process (copy to GPU → kernel function execution → copy to host). Each step was strictly followed in the streams.

In the program, image is partitioned into K blocks.

For an image data block with index i （i∈{0..K-1}）, its original position is:

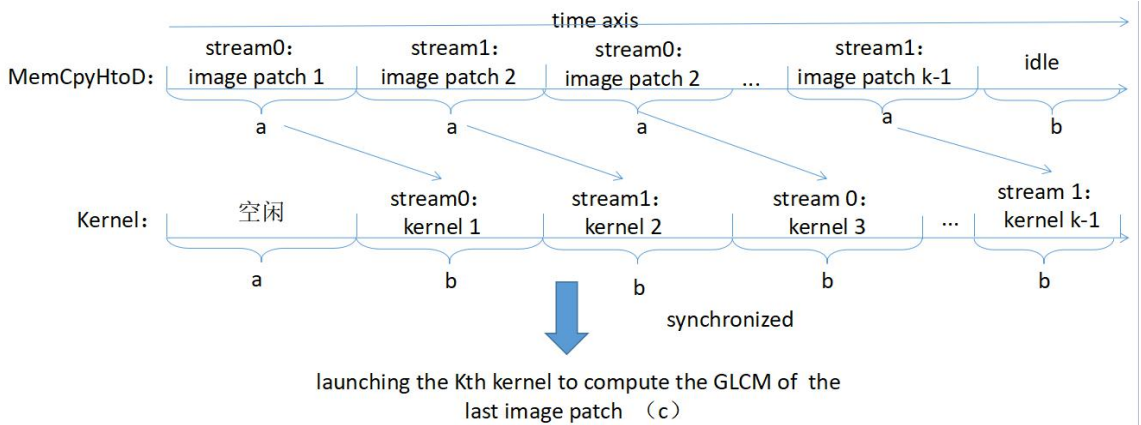

Fig.3 Diagram of asynchronous execution based on CUDA streams

As illustrated in Table 3, in the whole process (as shown in Fig.3, a represents the time of block data transmission to GPU and b represents the time of kernel processing block data), the time spent on the transmission from the host to device accounted for approximately 50% of the total time. Hence, data transmission is one of the major bottlenecks that affect the program performance. Next algorithm design and program optimization were conducted based on data memory access.

CUDA streams are a series of orders executed in sequence, but different streams can execute their own order at the same time or not regardless of sequence. As shown in Fig.3, in our realization, the CPU codes can run as the kernel initiates and the memory copies operate at the same time by the application of streams and pinned memory. In our realization, two CUDA streams were defined, i.e.,

$$offset\_start(i) = \frac{N \times N}{K} \times i \quad (7)$$

The position varies according to two conditions when it comes to the last index:

$$offset\_end(i) = \begin{cases} \frac{N \times N}{K} \times (i+1) + Pad, & i \leq K-1 \\ \frac{N \times N}{K} \times (i+1), & i = K \end{cases} \quad (8)$$

Where Pad:

$$Pad = \begin{cases} d, & \theta = 0°, i \leq K-1 \\ d \times (N-1), & \theta = 45°, i \leq K-1 \\ d \times N, & \theta = 90°, i \leq K-1 \\ d \times (N+1), & \theta = 135°, i \leq K-1 \end{cases} \quad (9)$$

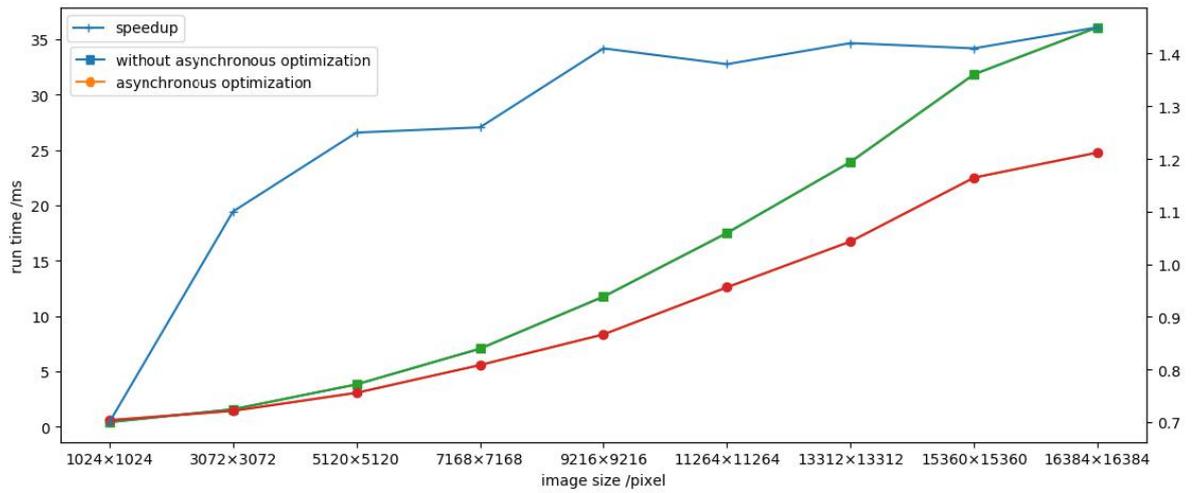

Fig.4 Comparison of asynchronous optimization efficiency

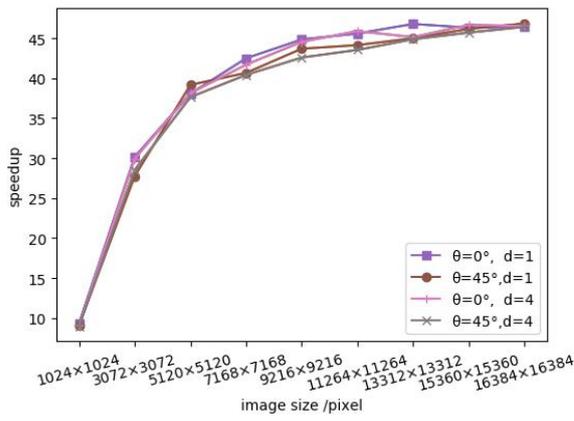

Fig.5(a)

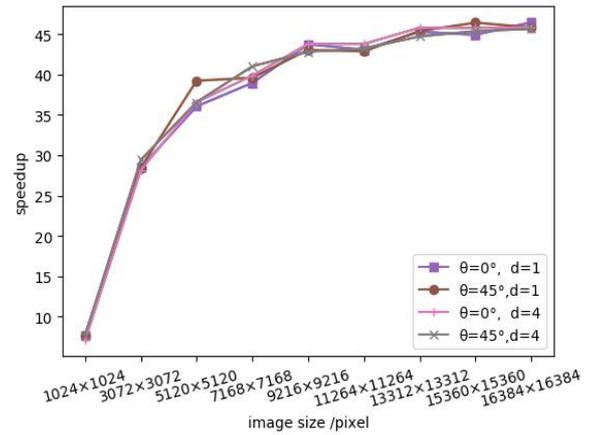

Fig.5(b)

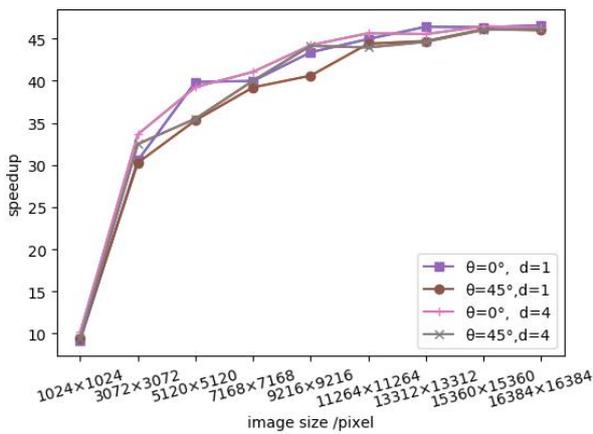

Fig5(c)

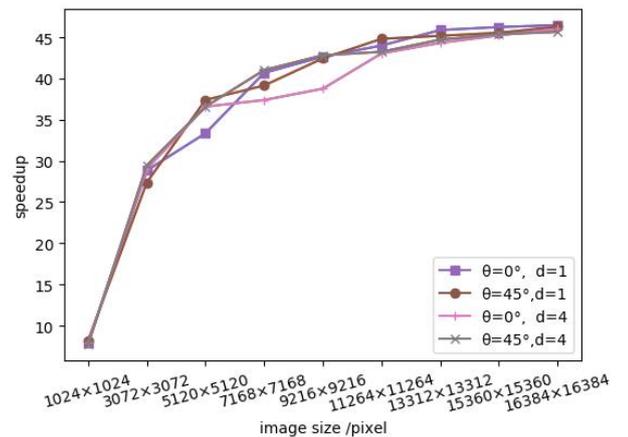

Fig5(d)

Fig.5 Compared to CPU, the speed-up ratio of GPU

When i ≤ K, Formula (8) and Formula (9) are feasible for the purpose of ensuring the pixel values of image blocks among [i×N×N/K -Pad ,.., i×N×N/K -Pad-1] can be counted and vote.

For the verification of the asynchronous efficiency, 10 groups of images of different resolutions were conducted based on the basic GPU and the asynchronous GPU (d=1, θ=45°, gray level of 32, Fig.1(a)). Results are shown in Fig.4.

With the resolution increase, the computation time in Scheme 2 and Scheme 3 raised as the speed-up ratio between the two schemes were expanding. Eventually, the speed growth stayed at the level of about 10%.

Scheme 3 is the final scheme for optimization. For the verification of the speed advantage of the algorithm in this study, a serial program to be executed on CPU based on C was written. The speed-up ratios between CPU and GPU were illustrated in Fig.5, taking the gray levels of 8 and 32, and the images in Fig.1(a) and Fig.1(b) as example. Compared to CPU, the speed-up ratio of GPU reached as high as 50 times, fully indicating the GPU advantage on the GLCM computation.

## IV. Conclusion

This paper analyzes the parallelism of gray level co-occurrence matrix, put forward to independence in the image of adjacent pixels on operation using each thread in CUDA, and perform the analysis and parallel processing methods of gray level co-occurrence matrix. It is proven that the calculation of gray level co-occurrence matrix based on GPU is feasible.

## V. Acknowledgment

This work is financially supported by Huaqiao University graduate research innovation ability cultivation fund (Grant no. 151322001), the Science and Technology Project of Xiamen (Grant no. 3502Z20173045),and the Science and Technology Project of Quanzhou (Grant no. 2016G051).